# Structure and dynamics of dynorphin peptide and its receptor

**Guillaume Ferré*, Georges Czaplicki, Pascal Demange, Alain Milon***

Institut de Pharmacologie et de Biologie Structurale (IPBS CNRS UPS), Université de Toulouse, 31077 Toulouse, France

* Corresponding authors: Email addresses: guillaume.ferre@ipbs.fr (G. Ferré); alain.milon@ipbs.fr (A. Milon)

## Abstract

Dynorphin is a neuropeptide involved in pain, addiction and mood regulation. It exerts its activity by binding to the kappa opioid receptor (KOP) which belongs to the large family of G-protein coupled receptors. The dynorphin peptide was discovered in 1975, while its receptor was cloned in 1993. This review will describe: *a)* the activities and physiological functions of dynorphin and its receptor, *b)* early structure-activity relationship studies performed before cloning of the receptor (mostly pharmacological and biophysical studies of peptide analogues), *c)* structure-activity relationship studies performed after cloning of the receptor via receptor mutagenesis and the development of recombinant receptor expression systems, *d)* structural biology of the opiate receptors culminating in X-ray structures of the four opioid receptors in their inactive state and structures of MOP and KOP receptors in their active state. X-ray and EM structures are combined with NMR data, which gives complementary insight into receptor and peptide dynamics. Molecular modelling greatly benefited from the availability of atomic resolution 3D structures of receptor-ligand complexes and an example of the strategy used to model a dynorphin-KOP receptor complex using NMR data will be described. These achievements have led to a better understanding of the complex dynamics of KOP receptor activation and to the development of new ligands and drugs.

## I. Dynorphin: a neuropeptide involved in pain, addiction and mood regulation

Dynorphin is an endogenous neuropeptide first isolated from porcine pituitary (Cox, Opheim, Teachemacher, & Goldstein, 1975) with a particularly potent opioid activity (Goldstein, Tachibana, Lowney, Hunkapiller, & Hood, 1979). The dynorphin A 1-17 isoform was the first to be fully sequenced (Goldstein, Fischli, Lowney, Hunkapiller, & Hood, 1981) and revealed an amino-terminal sequence identical to leu-enkephalin opioid peptide (YGGFL) with a basic carboxy-terminal extension. Several dynorphin isoforms were further identified: dynorphin A 1-8, dynorphin B 1-13, big dynorphin and leumorphin (Charles Chavkin, 2013) (**Fig. 1**). All dynorphin isoforms and α- and β-neo-endorphin, which are also leu-enkephalin-based opioid peptides, derive from a common precursor named prodynorphin (C. Chavkin, Bakhit, Weber, & Bloom, 1983; Kakidani et al., 1982; Weber, Evans, & Barchas, 1982; Zamir, Palkovits, Weber, MEzey, & Brownstein, 1984) while other opioids are derived from the precursors proenkephalin and proopiomelanocortin (Charles Chavkin, 2013). Prodynorphin and its processing enzyme, prohormone convertase 2 (PC2), are largely expressed in the central nervous system (Berman et al., 2000; Civelli, Douglass, Goldstein, & Herbert, 1985) and dynorphin peptides are present in presynaptic neurosecretory vesicles (Molineaux & Cox, 1982; Pickel, Chan, & Sesack, 1993; Pickel, Chan, Veznedaroglu, & Milner, 1995; Whitnall, Gainer, Cox, & Molineaux, 1983). They can be released by membrane depolarization (C. Chavkin, et al., 1983) and subsequently activate the kappa opioid receptor (KOP) (Wagner, Evans, & Chavkin, 1991) which modulates neurotransmitter release and postsynaptic neural activity (Wagner, Terman, & Chavkin, 1993).

KOP belongs to the opioid receptor family which is composed of several subtypes, originally defined by the pharmacological profiles of the first receptors to be characterized (Dhawan et al., 1996): the μ and δ opioid receptors (MOP and DOP). KOP was initially named from ketocyclazocine opioid activity (Martin, Eades, Thompson, Huppler, & Gilbert, 1976) and highly KOP-specific agonists were synthesized (Dhawan, et al., 1996) such as U50488 (Lahti, VonVoigtlander, & Barsuhn, 1982) or U69593 (Lahti, Mickelson, McCall, & Von Voigtlander, 1985). Opioid receptor cloning led to the identification of a fourth member of the family, the nociceptin opioid receptor (NOP), and of its endogenous ligand, nociceptin (J-C. Meunier et al., 1995; Mollereau et al., 1994). Several studies reported the cloning of KOP from rodents (Li et al., 1993; Meng et al., 1993; Yasuda et al., 1993) and then from human (Simonin et al., 1995; J. Zhu et al., 1995), showing that it, together with the other opioid receptors, belongs to the G-protein-coupled receptors (GPCR) superfamily. Opioid receptors are mainly $G_i/G_o$-coupled (Al-Hasani & Bruchas, 2011; Prather et al., 1995). They display a small basal intracellular signaling activity, at about 10% of the maximal response for KOP, in the absence of any ligand, and this is modulated by extracellular ligand binding (D. Wang, Sun, & Sadee, 2007).

| Peptide | Sequence | pKi KOP | pKi MOP | pKi DOP | References |
|---|---|---|---|---|---|
| Met-enkephalin | YGGFM | 7.3 | 8.7 | 9.3 | 1, 2, 3 |
| Leu-enkephalin | YGGFL | 7.0 | 8.2 | 9.4 | 1, 4 |
| Dynorphin A 1-8 | YGGFLRRI | 9.3 | 8.6 | 8.9 | 4, 5, 6 |
| Dynorphin A 1-13 | YGGFLRRIRPKLK | 9.6 | 8.4 | 8.3 | 4 |
| Dynorphin A 1-17 | YGGFLRRIRPKLKWDNQ | 9.4 | 8.6 | 8.9 | 4, 5, 6, 7, 8 |
| Big dynorphin | YGGFLRRIRPKLKWDNQKRYGGFLRRQFKVVT | 8.5 | 8.2 | 8.5 | 6 |
| Leumorphin | YGGFLRRQFKVVTRSQQDPNPNAYYGGLFNV | 9.3 | 8.4 | 8.4 | |
| Dynorpin B 1-13 | YGGFLRRQFKVVT | 9.3 | 8.7 | 8.5 | 6 |
| α-neo-endorphin | YGGFLRKYPK | 9.9 | 8.9 | 8.9 | 3, 5, 7 |
| β-neo-endorphin | YGGFLRKYP | 9.3 | 8.3 | 8.9 | 5 |

**Figure 1:** Dynorphin-related opioid peptides. Met-enkephalin, leu-enkephalin, dynorphin A 1-13 and prodynorphin-derived peptides amino acid sequence and affinity for opioid receptors. The common N-terminal leu-enkephalin "message" sequence is colored in green and C-terminal "address" residues conserved with dynorphin A 1-17 in blue. In order to compare affinities in homogenous systems, $pK_i$ values from (Mansour, Hoversten, Taylor, Watson, & Akil, 1995) are reported where competition binding experiments were conducted against rat opioid receptors transiently expressed in COS-1 cells. Binding properties have been additionally reviewed from the indicated references from IUPHAR/BPS guide to pharmacology (Alexander et al., 2017): 1 (Meng, et al., 1993), 2 (Raynor et al., 1994), 3 (Yasuda, et al., 1993), 4 (Toll et al., 1998), 5 (Simonin, et al., 1995), 6 (Merg et al., 2006), 7 (J. Zhu, et al., 1995), 8 (J. Zhu, Luo, Li, Chen, & Liu-Chen, 1997). Similar trends are observed despite discrepancies arising from differences in experimental models and conditions.

Opioid receptor signaling controls a multitude of intracellular effectors by both G-protein-dependent and G-protein-independent pathways (Al-Hasani & Bruchas, 2011; Bruchas &

Chavkin, 2010; Law, Wong, & Loh, 2000). G-protein-dependent opioid receptor signaling predominantly results in neuronal excitability and synaptic transmission inhibition. $G_i/G_o$-protein activation promotes G-protein-coupled inwardly rectifying potassium channel (GIRK) activity (Barchfeld & Medzihradsky, 1984; Childers & Snyder, 1978; Minneman & Iversen, 1976). It causes neuronal membrane hyperpolarization and thus attenuates the neuron's ability to generate and propagate action potentials (Henry, Grandy, Lester, Davidson, & Chavkin, 1995; Sadja, Alagem, & Reuveny, 2003; Schneider, Eckert, & Light, 1998). Activation of $G_i/G_o$-proteins also inhibits voltage-dependent calcium channels, diminishing calcium influx in response to action potentials, and thus prevents neurotransmitter synaptic release (Bourinet, Soong, Stea, & Snutch, 1996; Rhim & Miller, 1994; Rusin, Giovannucci, Stuenkel, & Moises, 1997; Zamponi & Snutch, 1998). Activated $G_i/G_o$-proteins also inhibit adenylate cyclase thus decreasing intracellular cyclic adenosine monophosphate (cAMP) concentration (Taussig, Iniguez-Lluhi, & Gilman, 1993), which in turn regulates numerous targets. Opioid receptors also signal through G-protein-independent pathways. Following agonist stimulation, they can be intracellularly phosphorylated by G protein-coupled receptor kinases (GRKs) leading to receptor interaction with β-arrestins (Al-Hasani & Bruchas, 2011; Law, et al., 2000). This interaction is involved in receptor internalization, one of the consequences of which is to prevent exposition to extracellular ligands. It is also responsible for intracellular signaling *per se*, notably through mitogen activated kinase (MAPK) pathways that regulate predominant cellular processes such as gene transcription, proliferation and differentiation (Bruchas & Chavkin, 2010; Raman, Chen, & Cobb, 2007).

Because of their involvement in cellular communication, opioid peptides and their receptors participate in numerous physiological processes (Y. Feng et al., 2012). First, opioid receptor activation results in spinal and supra-spinal pain modulation (Ahlbeck, 2011; Waldhoer, Bartlett, & Whistler, 2004). The afferent nociceptive influx from peripheral tissues is regulated by descending neurons and opioid receptors participate in this system at multiple levels (Vanderah, 2010). As an example, KOP has been found to be present in spinal cord dorsal root ganglia (Attali & Vogel, 1989; Corder, Castro, Bruchas, & Scherrer, 2018; Ji et al., 1995) where it inhibits synaptic transmission between primary and secondary afferent neurons (Vanderah, 2010) in response to agonists released by descending neurons. Opioid peptides and their receptors also regulate monoaminergic systems and notably mesolimbic dopaminergic functions (Lutz & Kieffer, 2013). KOP agonists inhibit dopamine neuron release activity by their receptor-specific action both in the *nucleus accumbens* and the ventral tegmental area (Lutz & Kieffer, 2013; Margolis, Hjelmstad, Bonci, & Fields, 2003; R. Spanagel, A. Herz, & Shippenberg, 1992), resulting in an aversive effect (R. Bals-Kubik, A. Ableitner, A. Herz, & Shippenberg, 1993). In contrast, MOP agonists cause a rewarding effect (R. Bals-Kubik, et al., 1993) because they indirectly stimulate dopamine release by diminishing inhibitory gamma-aminobutyric acid (GABA) neuron activity in the ventral tegmental area (Lutz & Kieffer, 2013; R. Spanagel, et al., 1992). Opioid receptors thus mediate abusive opioid seeking behavior but are also more generally involved in modulating drug addictions with pronounced subtype discrepancies (Kreek et al., 2012). In addition, the mesolimbic dopamine reward system is linked to mood disorders (Nestler & Carlezon, 2006) and opioid receptors further regulate serotonin and noradrenaline neurons (Lutz & Kieffer, 2013), thought to participate in depression (Krishnan & Nestler, 2010). Consistent with this, opioid receptors and their ligands are involved in depression-like behaviours and KOP antagonists such as JDTic induce antidepressant-like effects (Lutz & Kieffer, 2013). Furthermore, opioid peptides and their receptors, especially dynorphin and KOP, control the hypothalamic–pituitary–adrenal axis (HPA) and are thus involved in stress-related phenomenon with implications in addiction, depression and anxiety disorders (Bruchas, Land, & Chavkin, 2010; Knoll & Carlezon, 2010; Kreek, et al., 2012; Ribeiro, Kennedy, Smith, Stohler, & Zubieta, 2005). Beside their predominant role in pain, addiction and mood control, opioid receptors have been linked to a multitude of functions such as immunity, neuroprotection, cell proliferation, neural differentiation, cardiovascular system regulation and feeding (Y. Feng, et al., 2012). Of particular importance, opioid receptors signaling stimulation can influence the respiratory system function (Pattinson, 2008) with strong respiratory depression observed upon activation of MOP (but not KOP). Together with its impact on breathing, MOP agonism induces an important inhibition of gastrointestinal transit (A. Tavani, P. Petrillo, A. La Regina, & Sbacchi, 1990), which represents a significant limitation of opioid use in the clinic.

The biological functions of opioid peptides and their receptors renders them key targets to interfere in pain, addiction and mood disorders. There is considerable interest in discovering new opioids with reduced side effects and compounds that target the dynorphin / KOP system are being developed in order to produce analgesic, antidepressant, anxiolytic or anti-addiction drugs (Dogra & Yadav, 2015; Zheng et al., 2017). Because of the biological significance of this system, extensive research is under way to decipher the molecular mechanisms underlying the dynorphin

/ KOP interaction and the resulting modulation of receptor signaling activity. These studies, which form the subject of this review, could also help in the design of novel KOP ligands with pharmacologically relevant properties such as biased agonism or allosteric modulation.

## II. The molecular mechanism of action of dynorphin: research and hypotheses prior to KOP cloning

The ability to produce dynorphin and analogues by solid-phase peptide synthesis (Goldstein, et al., 1979) opened the way to a wide range of structure-activity relationship studies on various peptide structures. These have been reviewed thoroughly elsewhere (Aldrich & McLaughlin, 2009; Lapalu et al., 1997; Naqvi, Haq, & Mathur, 1998; Ramos-Colon et al., 2016). In brief, the 17 amino-acid long dynorphin 1-17 may be shortened at its C-terminus to dynorphin 1-13 without affecting its activity (activation of KOP is typically assessed on the guinea pig ileum electrical contraction assay), and further to dynorphin 1-8 with a 50-fold reduction in activity and no loss in affinity (**Fig. 1** and (Mansour, et al., 1995)). Further shortening is deleterious for both affinity and activity. Substitution of Gly2 by L-amino-acids resulted in a reduction in activity while substitution by D-amino-acids resulted in reduced selectivity for KOP versus MOP, since D amino-acids tend to increase activity on MOP and to decrease it on KOP. Substitution of Pro10 by a D-proline led to a marked increase in selectivity for KOP over the other opiate receptors. The further alkylation of the amino group of Tyr1 gave rise to highly KOP-selective peptide ligands such as N–Benzyl[D–Pro10]–Dyn A(1–11) (Choi, Murray, DeLander, Caldwell, & Aldrich, 1992).

Peptide synthesis also enabled biophysical studies of dynorphin conformation and dynamics in solution and in membrane biomimetic media (Lancaster et al., 1991; Lind, Graslund, & Maler, 2006; Naito & Nishimura, 2004; Spadaccini, Crescenzi, Picone, Tancredi, & Temussi, 1999). These included circular dichroism, Raman, FT-IR and NMR spectroscopies. In aqueous solution, dynorphin is largely disordered. It may adopt secondary structures in specific environments, such as a type I β-turn involving the first five residues in DMSO (Renugopalakrishnan, Rapaka, Huang, Moore, & Hutson, 1988) or a α-helical conformation from residue Gly3 to Arg9 when bound to dodecylphosphocholine detergent micelles (Kallick, 1993). Limiting the conformational space available to constrained peptide analogues increases receptor binding affinity and specificity (Naqvi, et al., 1998). However, without knowing the structure adopted by dynorphin upon KOP binding, it was difficult to derive a clear understanding of the binding mechanism from these results.

A detailed study of dynorphin and dynorphin analogues bound to lipid bilayers was performed in the early eighties. Using a combination of infrared attenuated total reflection spectroscopy and capacitance minimization, dynorphin was found to bind to POPC lipid bilayers by forming an α-helix from Tyr1 to Pro10, which inserts into the bilayer perpendicular to the bilayer plane (Erne, Sargent, & Schwyzer, 1985). In the same study, the affinity for a neutral bilayer was determined to have a $K_d$ of 11 μM. In a subsequent theoretical estimation of the preferred orientation and binding energy of a series of dynorphin analogues of various lengths (from 1-13 to 1-5), a good correlation between the peptide amphiphilic moment, the affinity for lipid bilayers and the KOP receptor subtype specificity was found. Loss of the C-terminal positively-charged residues converts a KOP-selective peptide dynorphin, into a MOP/DOP-selective peptide Leu-enkephalin (**Fig. 1**). This, and similar observations on other neuropeptide families, gave rise to two major hypotheses: a) the *message-address concept*, first introduced in 1977 (Schwyzer, 1977)), in which the N-terminal 5 residues (message) are responsible for specific binding and receptor activation, while the positively charged C-terminus (address) is responsible for KOP receptor selectivity by concentrating the peptide in the vicinity of the receptor; b) the *membrane compartment concept*, in which the lipid bilayer plays an active role in catalyzing the peptide-receptor interaction (Auge, Bersch, Tropis, & Milon, 2000; Axelrod & Wang, 1994; Bersch, Koehl, Nakatani, Ourisson, & Milon, 1993; Czaplicki & Milon, 1998, 2005; Milon, Miyazawa, & Higashijima, 1990; Sargent & Schwyzer, 1986). One reason for the correlation between KOP receptor subtype specificity and the positively charged C-terminus (besides the potential role of the bilayer itself) became clear when the opiate receptors were cloned in 1992-1994 (Chen, Mestek, Liu, Hurley, & Yu, 1993; Evans, Keith, Morrison, Magendzo, & Edwards, 1992; Kieffer, Befort, Gaveriaux-Ruff, & Hirth, 1992; Mollereau, et al., 1994; Yasuda, et al., 1993). Indeed, it appeared that a specific feature of KOP and NOP receptors as compared to MOP and DOP receptors is the high negative potential surrounding the extracellular loops 2 (**Fig. 2**), which is expected to increase the local concentration of highly positively-charged neuropeptides (such as dynorphin and nociceptin for KOP and NOP, respectively). Dynorphin affinity for the extracellular loop 2 was further confirmed experimentally (Bjorneras et al., 2014). The sole effect of an attractive electrostatic potential may account for increased on-rate binding kinetics and receptor binding affinity by at least two orders of magnitude (Fersht, 1999).

```
            ECL1                  ECL2                    ECL3            ECL2
                                                                          charge
             -   +        + +-- -  -      ---    -    -
KOP: LMNSWPFGDVLCKI....GGTKVREDVDVIECSLQFPDDDYSWWD....EALGSTSHST-AALSSY   -6
                 +        + +      -             -    +      -
MOP: LMGTWPFGTILCKI....ATTKYRQ--GSIDCTLTFSHPTW-YWE....KALVTIPETT-FQTVSW    0
       -     -   +        + +-                   -        - -++-
DOP: LMETWPFGELLCKA....AVTRPRD--GAVVCMLQFPSPSW-YWD....WTLVDIDRRDPLVVAAL    0
                 +          --  -- -  -     -                 -
NOP: LLGFWPFGNALCKT....GSAQVED--EEIECLVEIPTPQD-YWG....QGLGVQPSSE-TAVAIL   -7
```

**Figure 2:** Primary sequences of the extracellular loops of the opioid receptor subtypes, KOP, MOP, DOP and NOP showing that the extracellular loop 2 (ECL2) is particularly rich in negative charges in KOP and NOP, while it is neutral for MOP and DOP. A similar trend is observed for the entire extracellular surface. This characteristic contributes to KOP specificity of positively charged dynorphin analogues as shown with MOP/KOP chimeric receptors (J. B. Wang, Johnson, Wu, Wang, & Uhl, 1994).

## III. KOP cloning, site-directed mutagenesis and heterologous expression systems

KOP (Li, et al., 1993; Meng, et al., 1993; Simonin, et al., 1995; Yasuda, et al., 1993; J. Zhu, et al., 1995) and the other three opioid receptors DOP, MOP and NOP (Chen, et al., 1993; Evans, et al., 1992; Kieffer, et al., 1992; Mollereau, et al., 1994; Yasuda, et al., 1993) were cloned in the early nineties. KOP shares 60% sequence similarity with the other receptors. The highest sequence diversity is located in the N-terminus, C-terminus and extracellular loops (Waldhoer, et al., 2004). Moreover, all four opioid receptors possess a remarkable signature, specific to class A GPCRs, which includes a sodium binding pocket (residues numbering corresponds to human KOP; superscripts are given according to Ballesteros-Weinstein numbering (Ballesteros & Weinstein, 1995)): D105$^{2.50}$, N141$^{3.35}$, S145$^{3.39}$ ; a PIF micro-switch P238$^{5.50}$, I146$^{3.40}$, F283$^{6.44}$ ; a NPxxY motif N326$^{7.49}$, Y330$^{7.53}$ ; a DRY motif D155$^{3.49}$, R156$^{3.50}$, Y157$^{3.51}$ and a cysteine disulfide bridge C131$^{3.25}$, C210ECL2 (**Fig. 3**). Heterologous expression and site-directed mutagenesis experiments were then used intensively to better understand the receptor architecture, ligand recognition specificity and activation mechanisms. It was shown that the extracellular loops play a major role in receptor subtype specificity (Metzger & Ferguson, 1995; Seki et al., 1998). Several key transmembrane residues were identified, such as $S187^{4.54}$ (Claude et al., 1996), $E297^{6.58}$ (Larson, Jones, Hjorth, Schwartz, & Portoghese, 2000; Sharma, Jones, Metzger, Ferguson, & Portoghese, 2001), $Y312^{7.35}$ (Metzger, Paterlini, Ferguson, & Portoghese, 2001) and $I316^{7.39}$ (Owens & Akil, 2002). Using KOP/DOP chimeric receptor, Kong et al. obtained indirect evidence that "agonists and antagonists bind to different domains of the cloned kappa opioid receptor" (Kong et al., 1994). However, recent structures of KOP in the presence of antagonist (H. Wu et al., 2012) and agonist (Che et al., 2018) do not support this statement, illustrating the difficulties of drawing firm conclusions from these approaches in the absence of precise 3D structures. Using the 3D structures of KOP, extensive mutagenesis experiments were performed to characterize the interaction of KOP with dynorphin, as well as with other non-peptide ligands such as salvinorin, which are summarized, together with previous studies, in Figure 3 and Table 1.

| KOP mutant | Binding | Potency | References |
|---|---|---|---|
| $T111^{2.56}$ | ↔ | ↓ | 3 |
| $Q115^{2.60}$ | ↓ | ↓ | 2 |
| $Y119^{2.64}$ | ↓ | ↓ | 1, 2 |
| $D138^{3.32}$ | ↓ | ↓ | 2, 3 |
| $Y139^{3.33}$ | ↔ | ↓ | 1, 2, 3 |
| $M142^{3.36}$ | ↔ | ↓ | 2 |
| $W287^{6.48}$ | ↔ | ↓ | 3 |
| $H291^{6.52}$ | ↓ | ↓ | 2, 3 |
| $I294^{6.55}$ | ↓ | ↓ | 2 |
| $Y312^{7.34}$ | ↔ | ↓ | 1, 2, 3 |
| $Y313^{7.35}$ | ↓ | ↓ | 1, 3 |
| $I316^{7.39}$ | ↓ | ↓ | 2 |
| $G319^{7.42}$ | ↔ | ↓ | 3 |
| $Y320^{7.43}$ | ↓ | ↓ | 1, 2, 3 |

**Table 1:** Main point mutations of KOP and their consequences on dynorphin binding and functional activity (cAMP inhibition assay). References: 1 (Yan et al., 2005); 2 (Vardy, et al., 2013); 3 (Che, et al., 2018).

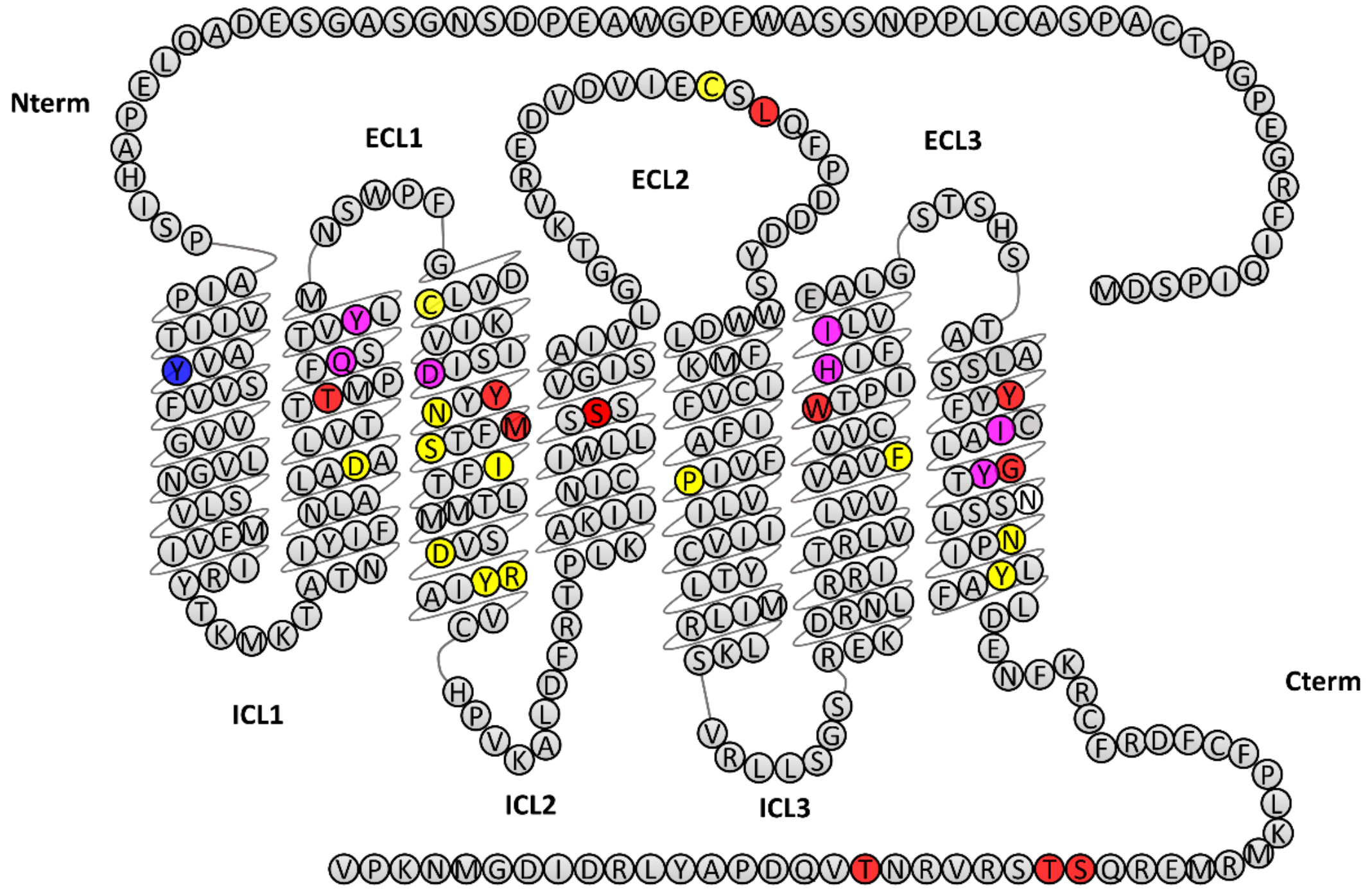

**Figure 3:** Primary sequence of human KOP highlighting mutations shown to affect KOP-dynorphin interactions. The limits of the seven α-helices were defined according to the activated KOP 3D structure (Che, et al., 2018). In the three-dimensional structure of inactive KOP (H. Wu, et al., 2012) these limits differ slightly for α-helix 5 (D218$^{5.30}$ to S260$^{5.72}$ instead of W221$^{5.33}$ to S255$^{5.67}$), α-helix 6 (R267$^{6.28}$ to L299$^{6.60}$ instead of R263$^{6.24}$ to G300$^{6.61}$) and α-helix 7 (L309$^{7.32}$ to L333$^{7.56}$ instead of T306$^{7.29}$ to D334$^{7.57}$). Mutations affecting dynorphin binding (more than 10-fold increase in $K_d$) are colored in blue (Y66$^{1.39}$), those affecting signaling (more than 10-fold increase in $EC_{50}$ in cAMP inhibition assay) are colored in red (T111$^{2.56}$, Y139$^{3.33}$, M142$^{3.36}$, S187$^{4.54}$, C210$^{ECL2}$, L212$^{ECL2}$, W287$^{6.48}$, Y312$^{7.35}$, G319$^{7.43}$, S356$^{Ct}$, T357$^{Ct}$, T363$^{Ct}$) and those affecting both are colored in magenta (Q115$^{2.60}$, Y119$^{2.64}$, D138$^{3.32}$, H291$^{6.52}$, I294$^{6.55}$, I316$^{7.39}$, Y320$^{7.43}$). Other residues generally considered to play a role in opioid receptor activation (Che, et al., 2018; Koehl et al., 2018) are colored in yellow (cysteine disulfide bridge C131$^{3.25}$, C210$^{ECL2}$; sodium binding pocket: D105$^{2.50}$, N141$^{3.35}$, S145$^{3.39}$; PIF microswitch: P238$^{5.50}$, I146$^{3.40}$, F283$^{6.44}$; NPxxY motif: N326$^{7.49}$, Y330$^{7.53}$; DRY motif: D155$^{3.49}$, R156$^{3.50}$, Y157$^{3.51}$). Residues D$^{3.32}$ and Y$^{7.43}$ (conserved in all four opioid receptors) were shown to form direct contacts with DAMGO N-terminus in the DAMGO-MOP-Gi structure (Koehl, et al., 2018) and thus presumably with the N-terminus of dynorphin in the case of KOP. Residue H$^{6.52}$ is conserved in MOP, KOP and DOP and forms a water-mediated contact with the phenol group of Tyr1 in the same MOP structure. Interestingly, this residue H$^{6.52}$ is mutated to a glutamine in NOP, whose ligand nociceptin possesses a phenylalanine at position 1 and is thus devoid of the phenol hydroxyl group. It should be noted that for clarity, this figure presents data obtained with dynorphin

(mostly from (Claude, et al., 1996, Che, 2018 #199; Vardy et al., 2013) and not with other non-peptide ligands for which other deleterious mutations have been described. Superscripts are given according to Ballesteros-Weinstein numbering (Ballesteros & Weinstein, 1995); residues numbering corresponds to human KOP.

## IV. Structure and dynamics of dynorphin and its receptor based on experimental 3D structures

Three-dimensional structures of GPCRs at atomic resolution began to appear with the structure of rhodopsin A (Palczewski et al., 2000). However, rhodopsin was a specific case due to its unusual stability and availability from natural sources, and further efforts were necessary to solve the three-dimensional structures of recombinant GPCRs. Several international consortia developed crucial methodologies in protein expression systems, receptor stabilization by mutagenesis, fusion proteins, the selection of stabilizing ligands, binding to antibodies (particularly nanobodies), the development of better solubilizing and crystallizing media (Caffrey & Cherezov, 2009; Cherezov, 2011; Granier & Kobilka, 2012; Kobilka & Schertler, 2008; Rosenbaum, Rasmussen, & Kobilka, 2009; Stevens et al., 2013; Tate, 2012).

Most GPCR structures were obtained from membrane proteins expressed in eukaryotic insect cells where the flexible N- and C-termini, as well as the intracellular loops (mostly ICL3), were deleted or replaced by exogenous protein domains promoting thermostability and crystallization, such as T4 lysozyme or the thermostabilized apocytochrome b562 RIL (BRIL) (Chun et al., 2012; Lv et al., 2016). Other expression systems have been used such as the methylotrophic yeast *P. pastoris* (Talmont, Sidobre, Demange, Milon, & Emorine, 1996), which allows stable isotope labelling, including perdeuteration (Massou et al., 1999), mostly for NMR experiments (Eddy et al., 2018). *E. coli* is also an interesting host for isotope-labelled GPCR biosynthesis which can be achieved by receptor expression as inclusion bodies followed by *in vitro* refolding during the protein purification (Baneres et al., 2003; Baneres, Popot, & Mouillac, 2011; Casiraghi et al., 2016). We have shown this strategy to be efficient for KOP: the dynorphin-KOP interaction was measured in our laboratory using KOP expressed in *E. coli* and refolded (unpublished results), and the same results were obtained as with KOP produced and purified from sf9 cell membranes (O'Connor et al., 2015).

These methodological developments allowed the first three-dimensional structure of a recombinant GPCR, the β2-adrenergic receptor, to be solved in 2007 (Rosenbaum et al., 2007). This marked the beginning of a new era of GPCR structural biology: according to the database GPCRdb, 270 structures of receptor-ligand complexes had been solved by September 2018, including 52 unique receptor complexes (http://gpcrdb.org/structure/statistics). In 2011 the first structure of the β2-adrenergic receptor in its active form (that is in the presence of an agonist and a G-protein or a nanobody mimicking the G-protein) was solved, thus revealing for the first time the atomic details of an activation mechanism of a GPCR (Rasmussen, Choi, et al., 2011; Rasmussen, DeVree, et al., 2011). The field of opioid receptors followed closely this revolution, with structures of the four opioid receptors solved in their inactive state in 2012 (Granier et al., 2012; Manglik et al., 2012; Thompson et al., 2012; H. X. Wu et al., 2012), and later in their active states for MOP (Huang et al., 2015; Koehl, et al., 2018) and KOP (Che, et al., 2018). An overlay of KOP in its inactive state, in complex with the antagonist JDTic (PDB 4DJH), and in its activated state, in complex with the agonist MP1104 and a nanobody mimicking G-protein (Nb39) (PDB 6B7S) illustrates the general mechanism of activation (**Fig. 4**): it is characterized by outward movements of transmembrane helix 6 (TM6) (by 10 Å) and ICL2 and inward movements of TM5 and TM7, leading to the creation of an intracellular pocket into which G-proteins can penetrate. These movements are associated with a contraction (10% reduction in volume of the ligand binding pocket) of the extracellular portion in the active-state KOP, with extracellular loop 2 (ECL2) and TM4 and TM6 moving closer to the receptor core (Che, et al., 2018). Both ligands, the antagonist JDTic and the agonist MP1104 bind at a similar location, with conserved contacts, in particular a salt bridge to $D138^{3.32}$ in TM3 and a water-mediated hydrogen bond with the backbone carbonyl oxygen of $K227^{5.39}$. Comparison of the active and inactive states of KOP indicates structural changes involving several residues of TM3, which are thus believed to be critical for coupling ligand-mediated changes in the orthosteric site and the transducer interface. This coupling is in part mediated by changes in a sodium binding pocket (formed by residues $D105^{2.50}$, $N141^{3.35}$ and $S145^{3.39}$) which acts as a negative allosteric modulator at opioid receptors (Fenalti et al., 2014; V. Katritch et al., 2014; Pasternak, Snowman, & Snyder, 1975).

With these data in hand, it may appear that the structural biology of opioid receptors and the molecular details of their activation mechanism are now well understood. This is not entirely true for several reasons: firstly, GPCRs in general and opioid receptors in particular must be understood in terms of interactions with other intracellular protein partners such as arrestins (Kang et al., 2015; Zhou et al., 2017), and with phospholipids and cholesterol within membrane domains (Dawaliby et al., 2016; Lagane et al., 2000; Meral et al., 2018; Pucadyil & Chattopadhyay, 2006; Xu et al., 2006), where they can form homo- and hetero-oligomers (Ferre et al., 2014; Jordan & Devi, 1999). Secondly, another important emerging characteristic of GPCRs is their extremely complex conformational landscape, within which the X-ray structures determined to date should be viewed as specific snapshots (Casiraghi et al., 2016; Deupi & Kobilka, 2010; Hilger, Masureel, & Kobilka, 2018). Thirdly, it is extremely difficult to solve three-dimensional structures of complexes of a GPCR with its natural peptide agonist. So far, the structure of DOP in complex with a non-natural peptide antagonist has been solved (Fenalti et al., 2015), and the structure of MOP in complex with DAMGO, a highly specific synthetic peptide agonist analogue, and Gi heterotrimeric protein has been solved at 3.5 Å resolution by cryo-electron microscopy (Koehl, et al., 2018).

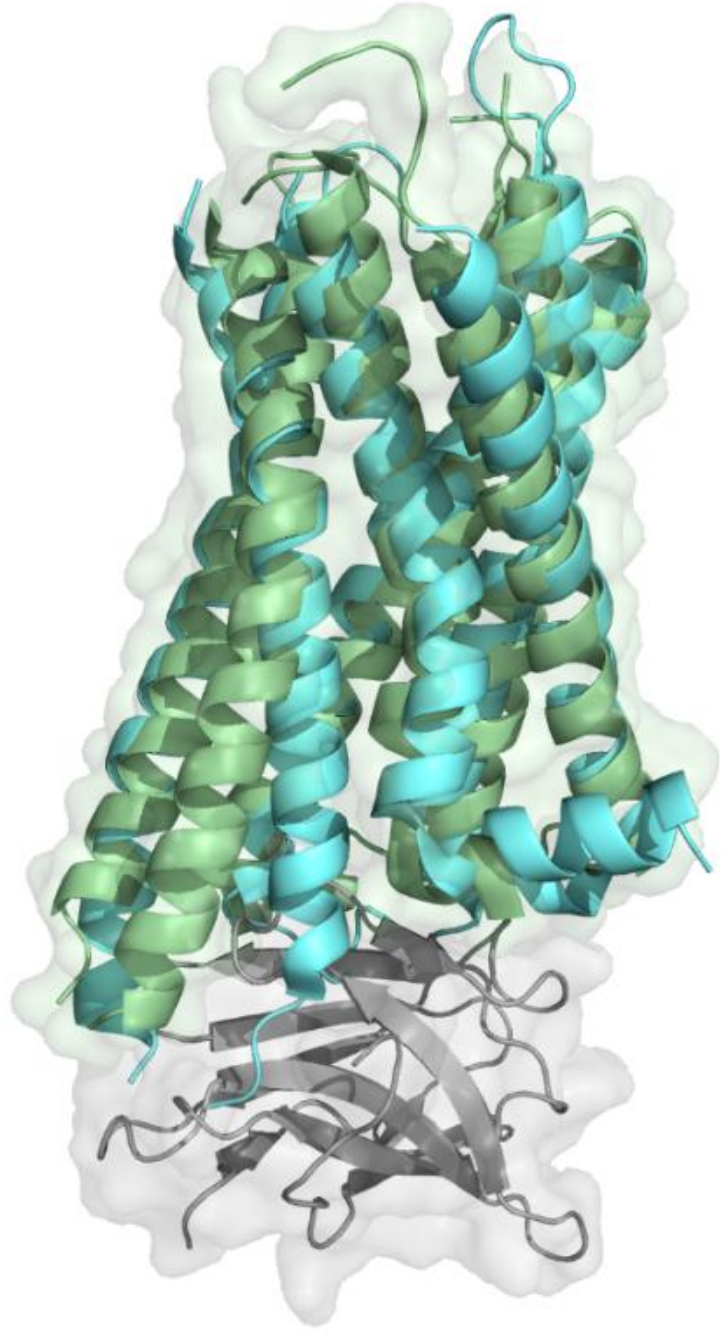

**Figure 4:** Superposition of the X-ray structures of KOP in its inactive state (in blue, PDB 4DJH) and in its active state (in green, PDB 6B7S). The nanobody present in the active state is shown in grey, penetrating a pocket created by the displacement of TM6. The antagonist and agonist in the inactive and active states respectively are not displayed.

NMR has proven highly efficient in demonstrating the conformational heterogeneity of GPCRs and their ligands (Bokoch et al., 2010; Casiraghi, et al., 2016; Didenko, Liu, Horst, Stevens, & Wuthrich, 2013; Nygaard et al., 2013), including opioid receptors (Sounier et al., 2015). In collaboration with R.C. Stevens and K. Wüthrich, we have solved the structure of dynorphin 1-13 bound to KOP in the absence of G-proteins by NMR, thus in its low affinity state for agonists (O'Connor et al., 2015). A well-defined α-helical conformation forms from Leu5 to Arg9 upon receptor binding (**Fig. 5A**). Most interestingly, $^{15}N$ relaxation measurements indicate that the peptide remains flexible on a nanosecond time scale in its receptor-bound state (**Fig. 5B**). This was expected for the C-terminus in which non-specific electrostatic interaction contribute to receptor binding (i.e. for the "address" part of dynorphin). It was however unexpected for the first four amino acids Tyr1-Gly2-Gly3-Phe4 which form the signature of opioid peptides (the "message") and which cannot be modified without affecting receptor binding and activation (Naqvi, et al., 1998). This mobility may be characteristic of an intermediate binding state observed for the G-protein uncoupled receptor, and work is in progress to determine whether N-terminal immobilization occurs in the high-affinity ternary complex of peptide, receptor and G-protein or the Nb39 nanobody used to stabilize the active conformation (Che, et al., 2018).

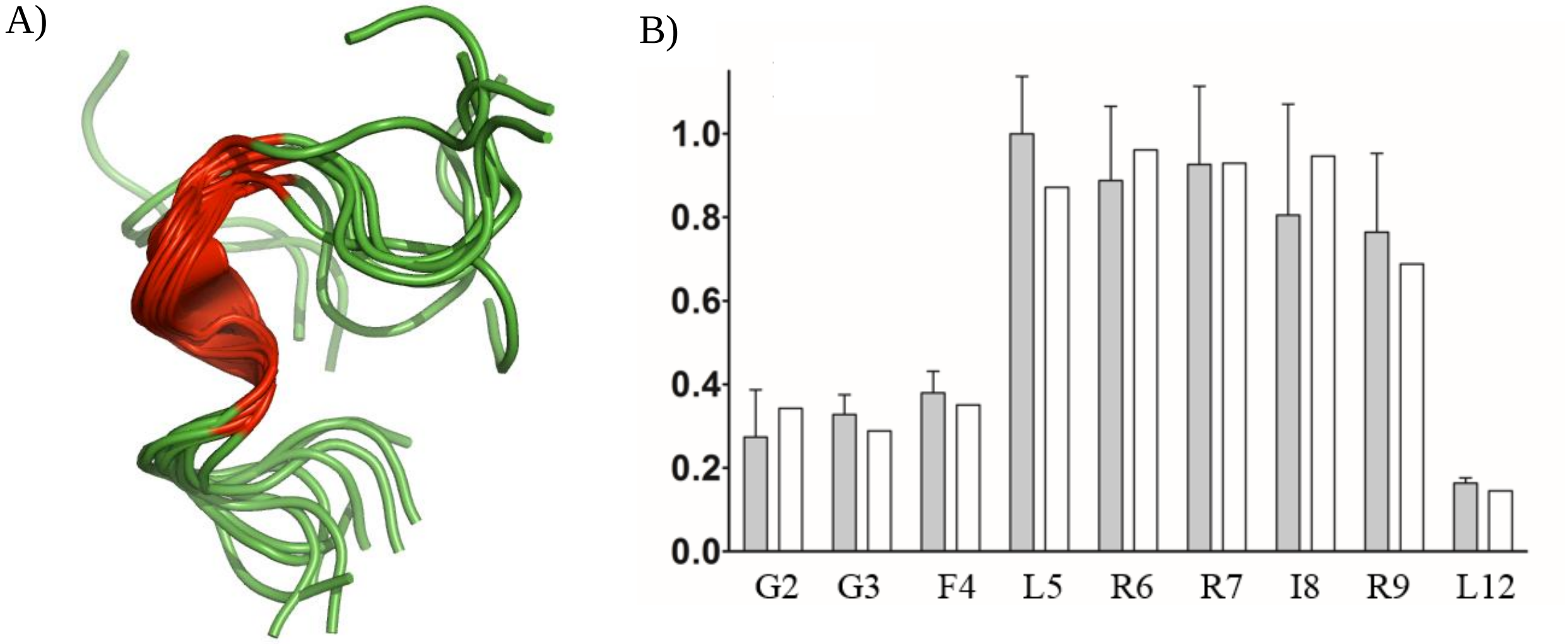

**Figure 5:** A) receptor-bound conformation of dynorphin 1-13: a α-helical conformation is formed between Phe4 and Arg9; B) Order parameter profile of dynorphin N-H bonds in the receptor-bound state. *Grey:* experimental data; *White:* calculated $S^2$ profiles from molecular dynamics simulations of dynorphin-receptor complexes. Note that both the N- and C-termini remain flexible in the receptor-bound state.

## V. Building 3D models of dynorphin-KOP complexes

Modelling structures of complexes formed by KOP and its agonists or antagonists has been attempted for more than 20 years (Alonso, Bliznyuk, & Gready, 2006; Bailey & Husbands, 2018; Benyhe, Zador, & Otvos, 2015; Bera, Marathe, Payghan, & Ghoshal, 2018; Gentilucci, Tolomelli, De Marco, & Artali, 2012; Johnson, 2017; Kane, Svensson, & Ferguson, 2006; Kaserer, Lantero, Schmidhammer, Spetea, & Schuster, 2016; Kolinski & Filipek, 2010; Lavecchia, Greco, Novellino, Vittorio, & Ronsisvalle, 2000; Martinez-Mayorga et al., 2013; Patra, Kumar, Pasha, & Chopra, 2012; Tessmer, Meyer, Hruby, & Kallick, 1997; Wu, Song, Graaf, & Stevens, 2017; Yongye & Martínez-Mayorga, 2012). Some of these models specifically focused on dynorphin (Bjorneras, et al., 2014; Charles Chavkin, 2013; Iadanza, Höltje, Ronsisvalle, & Höltje, 2002; Kang, et al., 2015; O'Connor, et al., 2015; Paterlini, Portoghese, & Ferguson, 1997; Sankararamakrishnan & Weinstein, 2000; Smeets et al., 2016; Vardy, et al., 2013). Early studies, performed before any experimentally determined receptor structures were available, were based entirely on modelling (Iadanza, et al., 2002; Paterlini, et al., 1997; Wan et al., 2000). Dynorphin was positioned within the receptor such that it was in agreement with mutagenesis data. Specifically, the spatial proximity between the N-terminus of dynorphin and residue $D138^{3.32}$ was preserved. Recent progress in obtaining X-ray and electron microscopy structures of the opioid receptors has enabled significant advances, due to the wealth of details for both the receptor structure and the binding modes of associated ligands. A binding mode was proposed for dynorphin 1-8, in which the peptide's N terminal mimics the orientation of the phenol-like ring of the JDTic antagonist (Vardy, et al., 2013). The structure of dynorphin 1-13 in the bound state was determined using transferred NOE experiments and that of the dynorphin-KOP complex was modelled using restraints from NMR, in particular N-H bond order parameters derived from $^{15}N$ $R_2$ relaxation rates, as discussed below (O'Connor, et al., 2015),.

A protocol has been developed for creating three-dimensional structures of dynorphin-KOP complexes (O'Connor, et al., 2015). The molecular modelling approach (**Fig. 6**) starts with the separate preparation of the two molecules. Although a KOP structure in complex with an antagonist was known from X-ray studies (H. Wu et al., 2012) a MOP structure mutated into KOP was used as starting point to avoid possible structural distortions caused by JDTic binding. Following the addition of missing residues and the prediction of rotamers of their side-chains (Krivov, Shapovalov, & Dunbrack, 2009; Nagata, Randall, & Baldi, 2012), the structure was relaxed and equilibrated in a 100 ns molecular dynamics (MD) run (R Salomon-Ferrer, Case, & Walker, 2013; Romelia Salomon-Ferrer, Götz, Poole, Le Grand, & Walker, 2013), followed by a clustering procedure which allowed major conformers of the receptor to be identified. The use of a cluster radius of 2 Å resulted in eight families of structures, whose representative members were selected as those being closest to the cluster centroids. Six major KOP structures were retained for docking. The modelling of dynorphin involved a typical MD simulation coupled with a simulated annealing protocol (Nilges, Clore, & Gronenborn, 1988), run in the presence of NMR constraints to preserve the peptide structure previously determined by NMR. The results indicated the existence of an α-helical turn involving residues Leu5-Ile8, and a disordered peptide elsewhere. In the subsequent docking procedure, the backbone of residues Leu5-Ile8 was therefore held fixed, while the rest of the molecule (backbone and side-chains) remained flexible. The KOP molecule was mostly held rigid, except for 16 residues whose side-chains were allowed to be flexible as they were considered likely to interact with the peptide in the binding pocket. Flexible docking was performed with the AutoDock Vina program (Trott & Olson, 2009), by launching multiple runs on each of the six retained KOP structures. The results were filtered to keep the 10 best poses per each KOP structure. This produced a set of 60 structures of the dynorphin-KOP complex, which was reduced to a set of 22 structures after selecting the best representatives from each family, characterized by lowest energies. The stability of the complexes was verified by running further MD simulations for times ranging from 50 to 100 ns.
The crucial final step consisted of comparing the values of the order parameters calculated from the generated structures with those obtained from NMR experiments. The calculations were based on the hypothesis that the flexible peptide adopts varying conformations and that its order parameter should be averaged over different conformers. The exchange is fast on the NMR time scale, but beyond the reach of MD simulations (1µs – 1ms). To find the minimum number of conformations required to reproduce the experimental order parameters, all combinations of modelled structures were taken into account. As a result, five major dynorphin conformations were identified, revealing significant structural diversity in both N- and C- termini.

Analysis of the resulting complexes allowed us to conclude that the position of a phenol-like functional group in the orthosteric site was largely conserved, for antagonist-bound structures of KOP-JDTic, as well as DOP-DIPP and MOP-funaltrexamine. One of the structural models (available in (O'Connor, et al., 2015)) revealed Tyr1 in a position near the phenol-piperidine fused-ring system of JDTic, resembling a previously proposed binding mode (Vardy, et al., 2013). Another model suggested that the side-chain of Tyr1 is close to the allosteric sodium binding site. $D138^{3.32}$ makes polar contacts with Tyr1, Gly2 and Gly3 in both of these models. However, only the latter features polar contacts of both Tyr1 and Arg7 with $W287^{6.48}$ and $N322^{7.45}$.

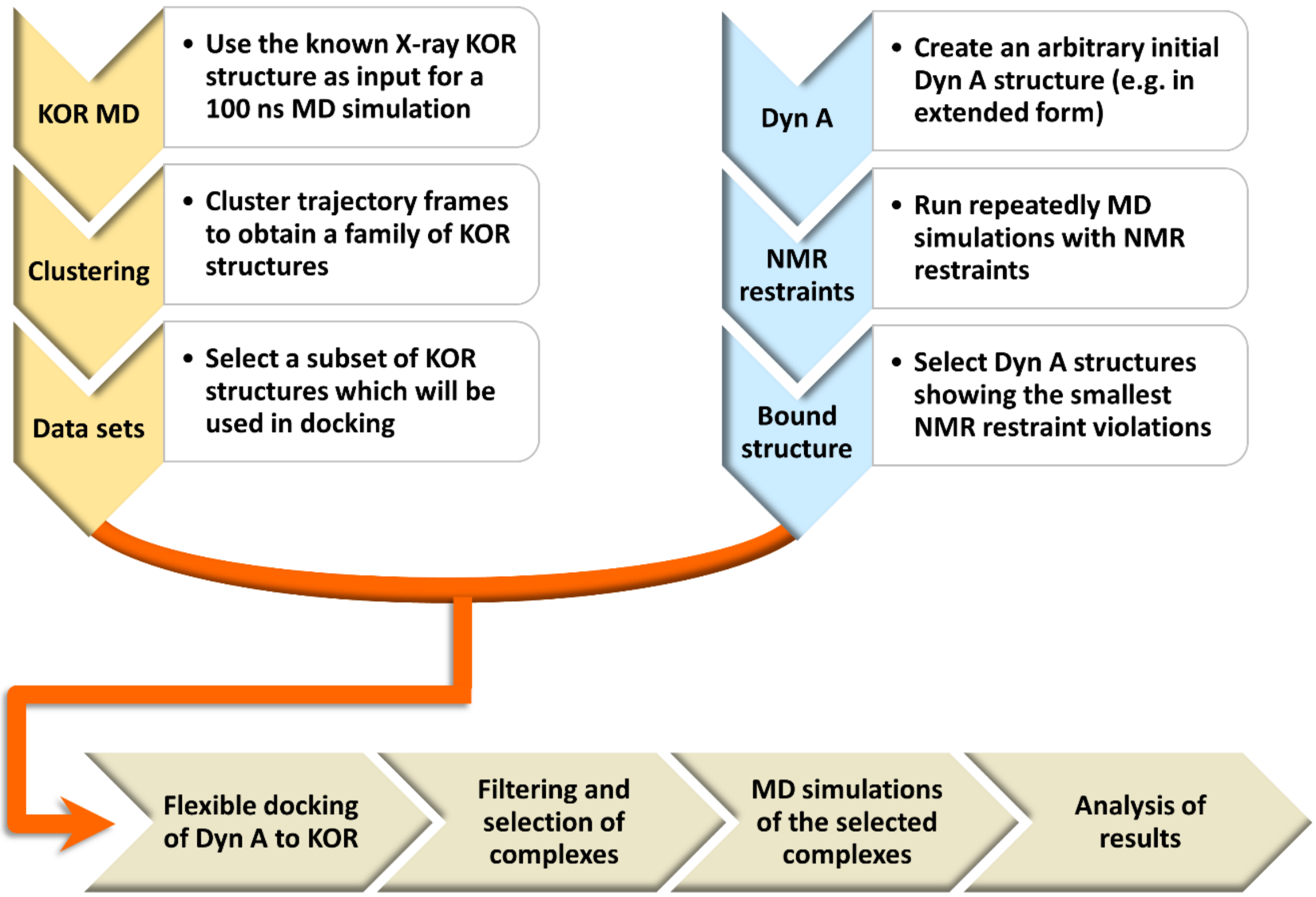

**Figure 6:** Molecular modelling protocol leading to three-dimensional structures of dynorphin-KOP complexes. Briefly, plausible KOP structures were obtained from MD simulations using an X-ray structure as input, while dynorphin models were obtained from simulations with NMR constraints. Subsequent docking, filtering and verification of stability of complex structures thus obtained permitted a selection of the optimal result. See the text for a more detailed description of the procedure.

## VI. Development of novel KOP ligands based on structural knowledge

Opioids possess powerful properties and are currently the most effective analgesics available, but development is still needed to reduce their undesired side-effects (Dogra & Yadav, 2015). KOP agonists produce analgesia but, for most, their use is limited by centrally-mediated adverse effects such as dysphoria. Nalfurafine (TRX-820), that does not induce dysphoria, has been registered in Japan since 2009 for the treatment of uremic pruritus (Kozono, Yoshitani, &

Nakano, 2018). Several peripherally-restricted (to avoid undesired effects) KOP agonists are currently in clinical trials: asimadoline (EMD-61753) (Delvaux et al., 2004; Szarka et al., 2007) and difelikefalin (CR845) (Hesselink, 2017) for irritable bowel syndrome and chronic or post-operative pain (https://clinicaltrials.gov/). Another promising strategy is the development of KOP-biased ligands: by acting as agonists for the G-protein-dependent signaling and not for G-protein-independent pathways, they could induce analgesia without dysphoria (Dogra & Yadav, 2015). In addition, there is a strong potential for KOP antagonists as antidepressant, anxiolytic or anti-addiction drugs (Zheng, et al., 2017). The knowledge obtained from the structures of inactive KOP (H. X. Wu, et al., 2012), KOP-bound dynorphin (O'Connor, et al., 2015) and G-protein bound KOP (Che, et al., 2018) offers new strategies for finding novel KOP ligands (Shang & Filizola, 2015).

Among the various ways of using structural knowledge in drug discovery, structure-based virtual screening has proven successful in designing GPCR ligands, such as for $D_3$ dopamine (Carlsson et al., 2011; Lane et al., 2013), $\beta_2$ adrenergic (Kolb et al., 2009; Weiss et al., 2013) and $A_{2A}$ adenosine (Carlsson et al., 2010; Vsevolod Katritch et al., 2010) receptors. It has allowed the discovery of a G-protein-biased MOP agonist that produces analgesia with diminished side-effects (Manglik et al., 2016). Studies using virtual screening from the inactive JDTic-bound KOP structure have discovered new KOP agonists (Negri et al., 2013) and new G-protein–biased agonists scaffolds (White et al., 2013). More recently, ligand-guided receptor optimization was applied to the inactive JDTic-bound KOP structure to generate alternative orthosteric pocket models that served to increase the prediction power of the methods (Zheng, et al., 2017). Virtual screening on the initial and alternative binding site models, followed by Tanimoto distances-based selection of novel chemotypes, revealed ligands in the micromolar range with a 32% hit rate. An initial optimization round generated eleven compounds with sub-micromolar affinities and functional assays defined two potent antagonists and one G-protein-biased agonist. The accumulation of precise structural knowledge on the modulation of KOP signaling activity by orthosteric ligands and allosteric modulators will certainly play a major role in future drug discovery programs.

## VII. Conclusions and future perspectives

Future research into the dynorphin-KOP structure and dynamics must address one major question: what defines a ligand as a full, partial, unbiased or biased agonist, an antagonist or an inverse agonist (Wacker, Stevens, & Roth, 2017). As GPCR signaling involves multiple receptor conformations in dynamic exchange, extensive research aims to characterize their conformational landscape and their modulation by ligands and signaling partners (Weis & Kobilka, 2018). Crystallography and electron microscopy provide structures of lowest-energy populated states (Wacker, et al., 2017). Spectroscopic methods such as NMR allow the characterization of dynamic properties (Weis & Kobilka, 2018), such as the weak allosteric coupling between the orthosteric site and the signaling interface, as described for MOP (Sounier, et al., 2015). One may take advantage of yeast or bacterial expression systems to produce specifically labelled GPCRs and perform advanced relaxation-based analyses to assess receptor dynamics, as was done recently for BLT2 (Casiraghi, et al., 2016) and A2A adenosine receptors (Clark et al., 2017; Eddy, et al., 2018). We are currently applying the methodologies we developed to determine the conformation and dynamics of KOP-bound dynorphin (O'Connor, et al., 2015) to the ternary dynorphin-KOP-Nb39 complex (where Nb39 is a nanobody which mimics G-proteins and confers high affinity binding to agonists). We thus hope to explain the 10-fold gain in dynorphin binding affinity in the presence of G-proteins. The conformational dynamics of G proteins and arrestins themselves can be modulated by the ligands (Hilger, et al., 2018). The question of conformational landscape and allosteric coupling must therefore be extended to entire GPCR signaling complexes and posed in the context of real cellular environment where modulation by lipids, membrane domains and other receptors do occur. While using information derived from structural biology, one should always bear in mind that recombinant G protein-coupled receptors *in vitro* may not recapitulate all the properties of native receptors naturally expressed in tissues as was shown for instance in (Broad et al., 2016).